\documentclass[preprint,aps,amsmath]{revtex4}
\usepackage{graphicx}
\usepackage{amsmath}
\usepackage{subfig}
\usepackage{bm}

\newcommand\bnabla{\boldsymbol \nabla}

\begin{document}
\preprint{}
\title{Asymmetric chiral alignment in magnetized plasma turbulence}
\author{A. Kendl}
\affiliation{Institute for Ion Physics and Applied Physics,  University of
  Innsbruck, A-6020 Innsbruck, Austria\vspace{2cm}
} 

\begin{abstract}
\vspace{0.5cm}
Multi species turbulence in inhomogeneous magnetised plasmas is found to
exhibit symmetry breaking in the dynamical alignment of a third species with
the fluctuating electron density and vorticity with respect to the magnetic
field direction and the species' relative background gradients. 
The possibility of truly chiral aggregation of charged molecules in magnetized
space plasma turbulence is discussed. 

\vspace{10cm}
This is a preprint version of a manuscript accepted 10/2012 for
publication in {\sl  Physics of Plasmas}. 
\end{abstract} 

\maketitle

\section{Introduction: drift wave turbulence}

Drift wave turbulence \cite{Hor99} is composed of coupled nonlinear
fluctuations of pressure $p$ and the electrostatic potential $\phi$ in an
inhomogeneous plasma with a guiding magnetic field ${\bf B}$. 
The local electrostatic potential acts as a stream
function for the dominant plasma flow with the drift velocity. 
The plasma is advected around the isocontours of a potential perturbation
perpendicular to the magnetic field and forms a quasi two-dimensional vortex:
an initial localized plasma density perturbation rapidly loses electrons along
the magnetic field ${\bf B}$, leaving a positive electrostatic potential
perturbation $\phi$ in its place. The associated 
electric field ${\bf E}  = - \bnabla \phi$ is directed outward from the
perturbation centre, and in combination with a magnetic field ${\bf B}$ causes
an "E-cross-B" vortex drift motion of the whole plasma along isocontours of
the potential with a velocity ${\bf v}_E = {\bf E} \times {\bf B}/B^2$. 

\bigskip

This gyration averaged drift motion is a result of the combined electric and
magnetic forces acting on a charged particle with small gyration radius
compared to background scales \cite{Hor99}. In the presence of a 
background density (or pressure) gradient $\bnabla n$ the vortex propagates in the
direction perpendicular to both $\bnabla n$ and ${\bf B}$.
Drift waves become unstable if the electron density can not rapidly adapt to
the electrostatic potential along the direction parallel to ${\bf B}$, which
can be caused by interaction of the electrons with other particles or
waves. Nonlinear coupling of drift waves results in a self-sustained fully
developed turbulent state \cite{Has83,Sco90}.  
 
\bigskip

The requirements for the occurrence of drift waves are ubiquitously met in
inhomogeneous magnetized plasmas when the drift scale $\rho_s = \sqrt{ (M_i T_e)/(eB)}$
is much smaller than the background pressure gradient length $L_p = (\bnabla \ln p)^{-1}$,
or equivalently, when the fluctuation frequency $\omega$ is much smaller than the ion
gyration frequency $\omega_{ci} = eB/M_i$. Here $M_i$ is the ion mass, $T_e$ the electron
temperature (in eV), $e$ the elementary charge, $B$ the magnetic field
strength, and $p = n_e T_e$ the electron pressure. 

\bigskip

Drift wave vortices are mainly excited in the size of a few drift scales $\rho_s$ and
nonlinearly dually cascade to both smaller and larger scales. This
perpendicular spatial scales of turbulent fluctuations are characteristically
much smaller than any parallel fluctuation gradients and scales, so that 
wave vectors fulfill  $k_{||} \ll  k_{\perp}$. 

\bigskip

Particle aggregation and transport in turbulence is a widely studied subject in fluid
dynamics \cite{Pro99}. In turbulent plasmas the aggregation of charged particles is in
addition influenced by static and dynamic electric and magnetic fields. For
gyrofluid drift wave turbulence in inhomogeneous magnetized plasmas it has been observed
by Scott~\cite{Sco05} that the gyrocenter density of light positive trace ions
(in addition to electrons and a main ion species) in three-species fusion
plasma drift wave turbulence tend to dynamically align with the fluctuating
electron density, so that fluctuation amplitudes of the electron and trace ion
densities are spatiotemporally closely correlated. 
Priego et al.~\cite{Pri05} have found that the particle density of inertial
impurties in a fluid drift wave turbulence model is also closely correlated
with the ExB flow vorticity.  

\bigskip

The present analysis shows that the dynamical alignment in drift wave
turbulence is not only sign selective with respect to the vorticity of trapping eddies,
for any given trace species charge and background magnetic field direction,
but also their large-scale background gradient. 

\bigskip

The resulting asymmetric effects on transport and aggregation of charged
particles is specifically discussed for application to space plasmas,
in particular for cases of a nearly collisionless plasma with cold
ions and low parallel electron resistivity, where the electrostatic potential closely
follows changes in the electron density. 
It is shown that in this case drift wave turbulence constitutes a unique
mechanism for truly chiral vortical aggregation of charged molecules in space
environments with a background magnetic field and plasma density gradients.

\bigskip

This paper is organized as follows: in section II the cold-ion gyrofluid model
equations for the present multi-species drift wave turbulence studies are
introduced. Section III describes the numerical methods, and in section IV the
computational results are presented. 
Section V discusses implications of the results for chiral particle
aggregation in turbulent space plasmas, which poses a possible
extraterrestrial physical  mechanism for achieving enantiomeric excess in
prebiotic molecular synthesis.   
The relation between the present gyrofluid model and the inertial fluid model
by Priego et al.~\cite{Pri05} is considered in the Appendix.

\bigskip

\section{Model: multi-species (cold ion) gyrofluid equations}

Multi-species drift wave turbulence in the weakly collisional limit can be
conveniently treated within a gyrofluid model.
Here a strongly reduced variant of the gyrofluid electromagnetic model
(``GEM'') by Scott~\cite{Sco10} is adapted 
for numerical computation of three-species drift wave turbulence in a quasi-2D
approximation in the electrostatic, isothermal, cold ion limit (neglecting
finite Larmor radius effects) with a dissipative parallel coupling model.  

The gyrocenter particle densities $n_s$ for (electron, main ion, and trace ion)
species $s \in (e,i,z)$ are evolved by nonlinear advection equations 
\begin{equation}
D_t n_s = (\partial_t + {\bf v}_E \cdot \bnabla) n_s = \partial_t n_s +
[\phi,n_s] = C_s, 
\label{e:conti}
\end{equation}
where all $C_s$  include hyperviscous dissipation terms $\nu_4 \nabla^4 n_s$
for numerical stability. The parallel gradient
of the parallel fluid velocity components are for simplicity  expressed by
assuming a single parallel wave vector $k_{||}$ in the force balance equation
along the magnetic field \cite{Has83}. Then $C_e$ in addition 
includes the coupling term $d(\phi - n_e)$, where $d$ is the parallel coupling
coefficient proportional to $k_{||}^2/\nu_{ei}$. For weakly dissipative
plasmas $d$ is well larger than unity, but for practical purposes $d = 2$
already sufficiently supports nearly adiabatic coupling between
$n_e$ and $\phi$ while allowing reasonable time steps. 
The coupling can be regarded as non-adiabatic when $d <1$.

The electrostatic potential is here derived from the local (linear) polarization equation 
\begin{equation}
\rho_m\nabla^2 \phi = n_e - n_i - a_z n_z,  
\label{e:locpol}
\end{equation}
where $\rho_m = 1+ a_z \mu_z$  with $a_z = Z n_z/ n_e$ and
$\mu_z = m_z/(Z m_i)$. In the local model only the density
perturbations are evolved in the advection equations, which then gain an
additional background advection term $g_s \partial_y \phi$ with $g_s = (L_{\perp}/L_{ns})$
on the right hand side, where $L_{\perp}$ is a normalising perpendicular scale
(here set identical to $L_{ne}$ so that $g_e=g_i=1$) and $y$ is the coordinate
perpendicular to both the magnetic field and the background gradient directions. 
In the following only trace ions with  $a_z \ll 1$ are considered, and the
parameter $a_z = 0.0001$ is going to be fixed, while other parameters are being varied.

The 2D isothermal gyrofluid model and its relation to (inertial)
fluid models is briefly discussed in the Appendix.

\section{Method: numerical simulation and analysis}

The numerical scheme for solution of the advection and polarization equations is
described (for a similar "Hasegawa Wakatani" type code where the ion density
equation is replaced by a vorticity equation) in refs.~\cite{Ken11,Shu11}. 

The present computations use a $512 \times 512$ grid corresponding to $(64\;\rho_s)^2$
in units of the drift scale. This is a resolution well established for drift
wave turbulence computations, resolving both the necessary small scales just below the gyro
radius (or drift scale), and also allowing enough large scales for fully
developed turbulent spectrum. The results converge for both higher
resolution and same size in drift units, and for larger size with accordingly
increased grid nodes.

The computations are initialized with quasi-turbulent random density fields
and run until a saturated state is achieved. 
The resulting turbulent advection 
dynamics is restrained to the 2-dimensional plane perpendicular to the (local)
magnetic field \cite{Ken08}.  
Statistical analysis is performed on spatial and temporal fluctuation data in this
saturated state. The turbulence characteristics of drift wave systems (like
power spectra, probability distributions, etc.) in general have been
extensively discussed before, so that here the focus is on analysis of the
correlation of the fluctuating density of the additional species with the
other fluctuating fields. 

\section{Results: asymmetric dynamical alignment}

In accordance with ref.~\cite{Sco05} we also observe in our simulations that 
the gyrocenter density of a low-density massive trace ion species tends to
dynamically align with the electron density $n_e$. 

The anomalous diffusion, clustering and pinch of impurities in plasma edge
turbulence has also been studied previously by Priego et al.~\cite{Pri05}.
In this paper the impurities were modeled as a passive fluid advected by the
electric and polarization drifts, while the ambient plasma turbulence was
modeled using the 2D Hasegawa-Wakatani model.
As a consequence of compressibility it has there also been found that the density
of inertial impurities correlates with the vorticity of the ExB velocity.
Trace impurities were observed to cluster in vortices of a precise orientation
determined by the charge of the impurity particles \cite{Pri05}. 

The major difference to the present approach is that Priego et al. have
considered only a constant impurity background (i.e. no impurity background
gradient), but have in addition included additional nonlinear inertial effects 
through the impurity polarization drift. Linear inertial effects by
polarization are already consistently included in the local gyrofluid model.
The detailed correspondence between the two models is discussed in the Appendix. 

For an initial homogeneous distribution of impurities with no background
gradient ($g_z=0$) and vanishing (or constant) initial perturbation $n_z(t=0)=0$ the
perturbed gyrocenter density $n_z(t)$ remains zero for all times, which
directly follows from the evolution equation $D_t n_z=0$. 
The perturbed fluid particle impurity density $N_z$ is related to the cold impurity gyrocenter
density by the transformation $N_z = n_z + \mu_z \Omega$ (see Appendix). 
As a result the perturbed fluid particle impurity density evolves according to
$D_t N_z = \mu_z D_t \Omega$ and directly follows the vorticity, which has
also been observed in fluid simulations by Priego et al.~\cite{Pri05}.
In the following the focus is on additional effects of a finite impurity
background density gradient specified by $g_z$.

In the adiabatic (weakly collisional) limit the electrons
can be assumed to follow a Boltzmann distribution with 
\begin{equation}
(n_0+n_e) = n_0 \exp[e\phi/T_e] \approx n_0 \; (1 + e\phi/T_e), 
\end{equation}
so that (for given $T_e$ and $n_0$) a
positive $n_e$ perturbation corresponds to a positive localized potential
perturbation $\phi$. The ${\bf v}_E \sim (-\bnabla \phi \times {\bf B})$ drift
motion of a plasma vortex around a localized $\phi$ perturbation possesses a
vorticity 
\begin{equation}
{\bf \Omega} = \bnabla \times {\bf v}_E = ({\bf B}/|B|) \nabla^2 \phi, 
\end{equation}
related to the  Laplacian of the electrostatic potential, and thus
(depending on the sign of the fluctuating potential) a definite sense of
rotation with respect to the background magnetic field ${\bf B}/|B|$. 

In ref.~\cite{Sco05} the absolute correlation $|r(n_e,n_z)|$ between the (gyrocenter)
density perturbations of electrons $n_e$ and trace ions $n_z$ has been
determined. Our present analysis in addition shows that under specific
conditions a definite sign relation appears for the sample correlation
coefficient 
\begin{equation}
r(n_e,n_z) =  { {\sum (n_e-\overline{n}_e)(n_z-\overline{n}_z) } \over
\sqrt{ \sum(n_e-\overline{n}_e)^2  \sum(n_z-\overline{n}_z)^2} }, 
\end{equation}
where the sum $\sum$ is taken over all grid points of the computational
domain, and the bar $\overline{n}_s$ denotes the domain average of the specific 
particle density. 

In local computations with the present model, where the densities are split
into a static spatially slowly varying background component $n_0$ with
perpendicular gradient lengths $L_n = (\nabla \ln n_0)^{-1}$ and a fluctuating
part with small amplitudes $n$ (typically in the range of a few percent), the
sign and value of $r(n_e,n_z) \approx \pm (0.90 \pm 0.02)$ are observed to
only depend on the relative sign but not the magnitude of the gradient lengths
$L_{ne}$ and $L_{nz}$, for a given direction of the background magnetic field
and fixed other parameters ($d=2$, $\mu_z = 10$). 

\begin{figure} 
\includegraphics[width=8.0cm]{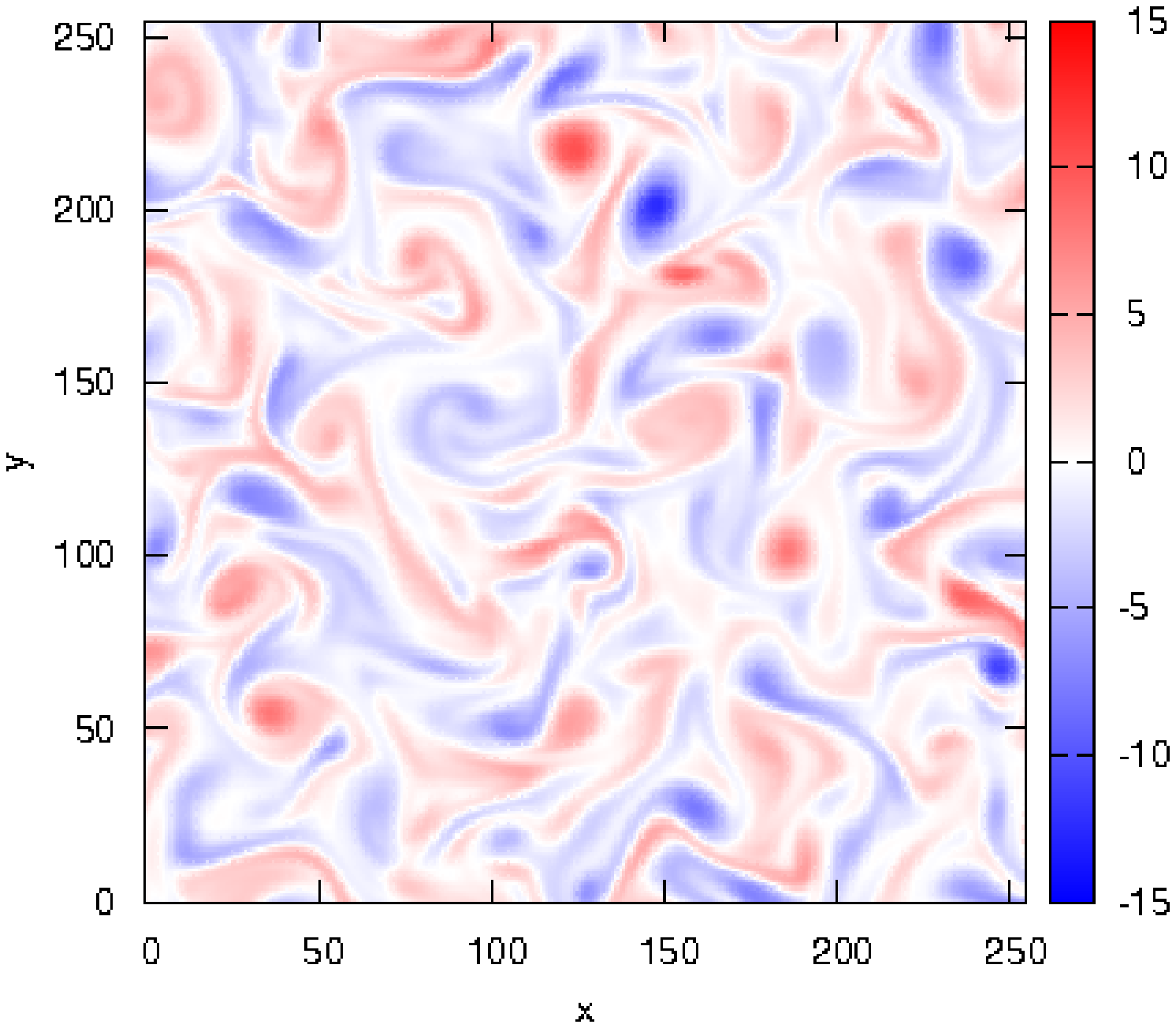}
\includegraphics[width=8.0cm]{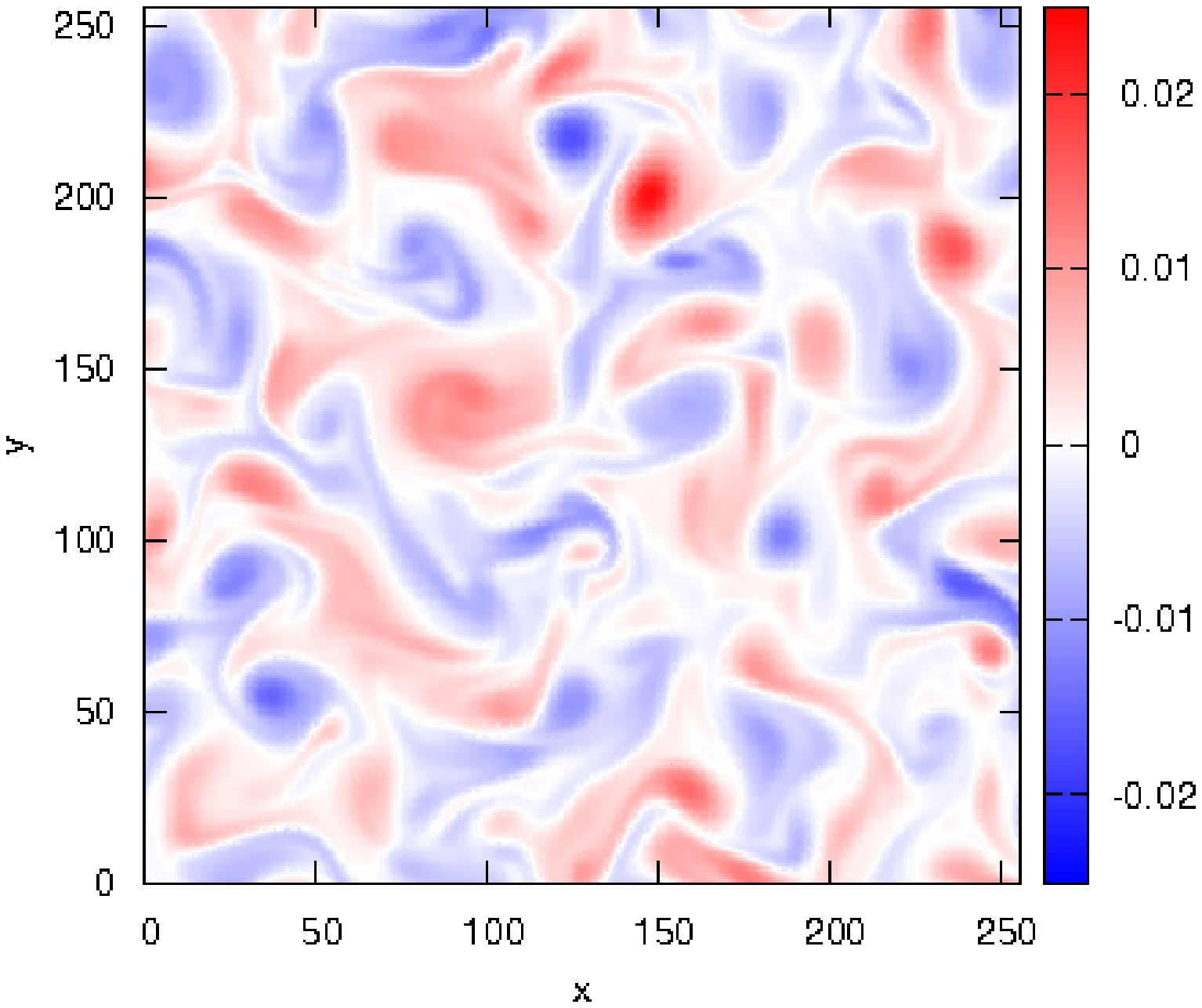}
\caption{ \sl
Left: Vorticity $\Omega(x,y)$; right: trace ion gyrocenter density
perturbation $n_z(x,y)$. 
Normalized amplitudes are used and only a  quarter of the complete
computational domain is shown. $\Omega$ and $n_z$ here show close negative alignment
for weak (quasi-adiabatic) coupling with $d=2$. The (co-aligned) impurity
gradient length is here set to $L_{\perp}/L_{nz} = + 0.001$.}
\label{f:fig1}
\end{figure}

The result for $r(n_e,n_z)$ changes only marginally for most other parameter
variations. In particular, the sign of the impurity charge $Z$ has no effect
on the alignment property. 
Stronger adiabaticity leaves $r (d=10) \approx \pm (0.90 \pm 0.02)$ largely
unchanged, while a smaller (strongly non-adiabatic) dissipative coupling coefficient
results in $r (d=0.01) \approx \pm (0.79 \pm 0.02)$. 
In Fig.~\ref{f:fig1} the computed turbulent fields of
vorticity (left figure) and a heavy (for example molecular) charged particle species
density perturbation (right) in the 2D plane perpendicular to {\bf B} are shown in a
snapshot to be closely (negatively) spatially correlated for parameters $d=2$
(quasi-adiabatic) and $g_z = L_{\perp}/L_{nz} = + 0.001$ (co-aligned
background density gradients).

\tabcolsep12pt
\begin{table} {Table 1: correlation coefficient $r(n_e,n_z)$}\\
\begin{tabular}{c|c|c} \hline
$g_z$ &  $d$ & $r(n_e,n_z)$ \\ \hline
 $+ 0.001$  & $2.00$  & $ + 0.90 \pm 0.01$ \\ 
 $+ 0.100$  & $2.00$  & $ + 0.90 \pm 0.01$ \\ 
 $+ 1.000$  & $2.00$  & $ + 0.91 \pm 0.01$ \\ 
 $- 1.000$  & $2.00$  & $ - 0.90 \pm 0.01$ \\ 
 $- 0.100$  & $2.00$  & $ - 0.91 \pm 0.01$ \\ 
 $- 0.001$  & $2.00$  & $ - 0.90 \pm 0.01$ \\ \hline

 $+ 0.001$  & $10.0$  & $ + 0.90 \pm 0.02$ \\ 
 $+ 0.001$  & $2.00$  & $ + 0.90 \pm 0.01$ \\ 
 $+ 0.001$  & $1.00$  & $ + 0.92 \pm 0.01$ \\ 
 $+ 0.001$  & $0.10$  & $ + 0.87 \pm 0.01$ \\ 
 $+ 0.001$  & $0.01$  & $ + 0.79 \pm 0.02$ \\ 
\hline
\end{tabular}
\end{table}

\bigskip

The observed sign-selective multi-species dynamical alignment
effect is basically caused by the (linear) drive of density fluctuations
$\partial_t n_z  \sim g_z \partial_y \phi$  by this background
advection term, where the potential $\phi$ is for each species just acting on
the respective background gradients with length $L_{ns}$. The species gyrocenter
densities are enhanced (positive partial time derivative) when the background
advection is positive, and decreased when negative. Co-aligned background gradients of
electrons and trace ions thus dynamically also imply co-alignment of the
gyrocenter density perturbations. In case of adiabatic electrons this additionally
implies negative alignment with vorticity, as can be seen in Fig.~\ref{f:fig1}. 
For counter-aligned background density gradients the plot for $n_z(x,y)$ in
Fig.~\ref{f:fig1} has exactly the same topology only with signs reversed,
i.e. blue and red in the colour scale exchanged.

\medskip

Strong non-adiabaticity ($d \ll 1$) only slightly reduces the
electron-to-impurity density correlation coefficient $r(n_e,n_z)$ compared to
adiabatic cases. Although the alignment between electron density and electrostatic
potential is reduced for a non-adiabatic response, the alignment between the
impurity gyrocenter density $n_z$ and vorticity $\Omega$ still remains.   

\bigskip

The results for $r(n_e,n_z)$ as a function of the dissipative coupling
parameter $d$ and (co- or counter-aligned) impurity gradient length $g_z =
L_{\perp}/L_{nz}$ are summarized in Table~1. 

\bigskip

\begin{figure} 
\includegraphics[width=8.0cm]{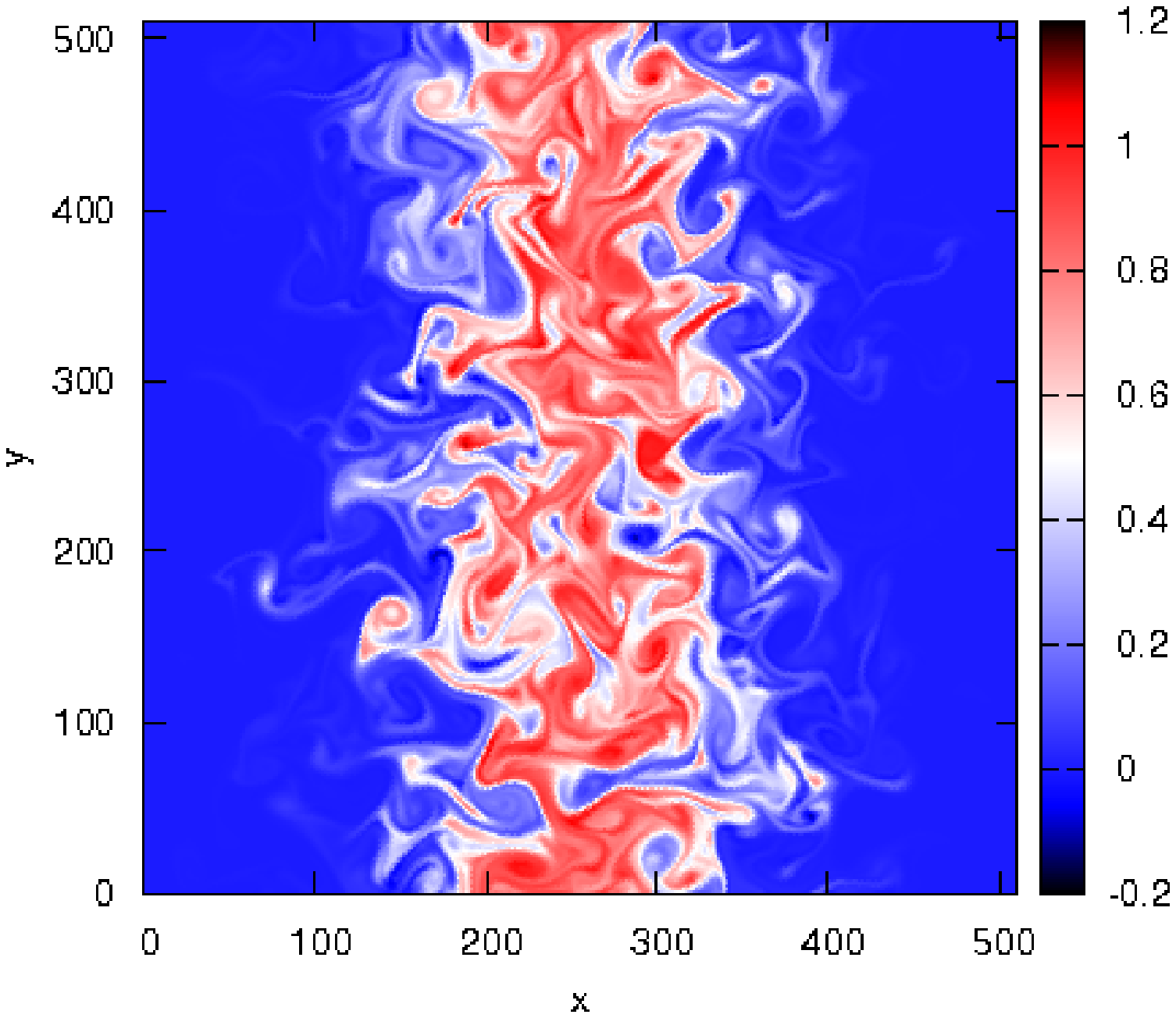}
\includegraphics[width=8.0cm]{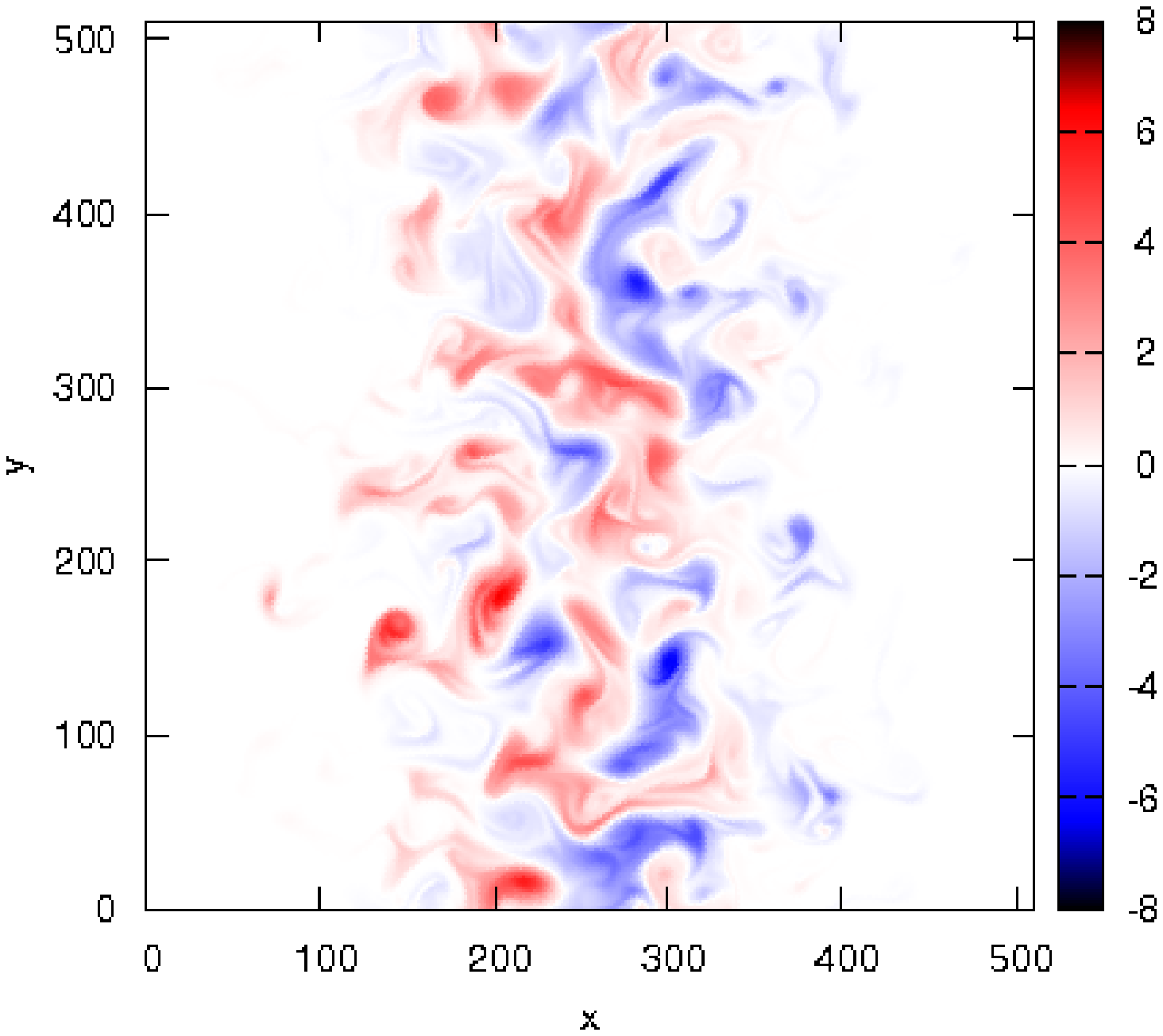}
\caption{ \sl
A Gaussian trace ion puff placed in the turbulent plasma is initially
localized around the center of the $x$ domain (and constant in $y$).
Left: impurity density $n_z(x,y)$. Right: the product function
$\Omega(x,y)\cdot n_z(x,y)$ shows in comparison with the top figure that
impurity density and vorticity are preferentially negatively aligned in the
right half of the domain (where the initial density gradients are co-aligned)
and positively aligned on the left. 
}
\label{f:fig2}
\end{figure}

\begin{figure} 
\includegraphics[width=8.0cm]{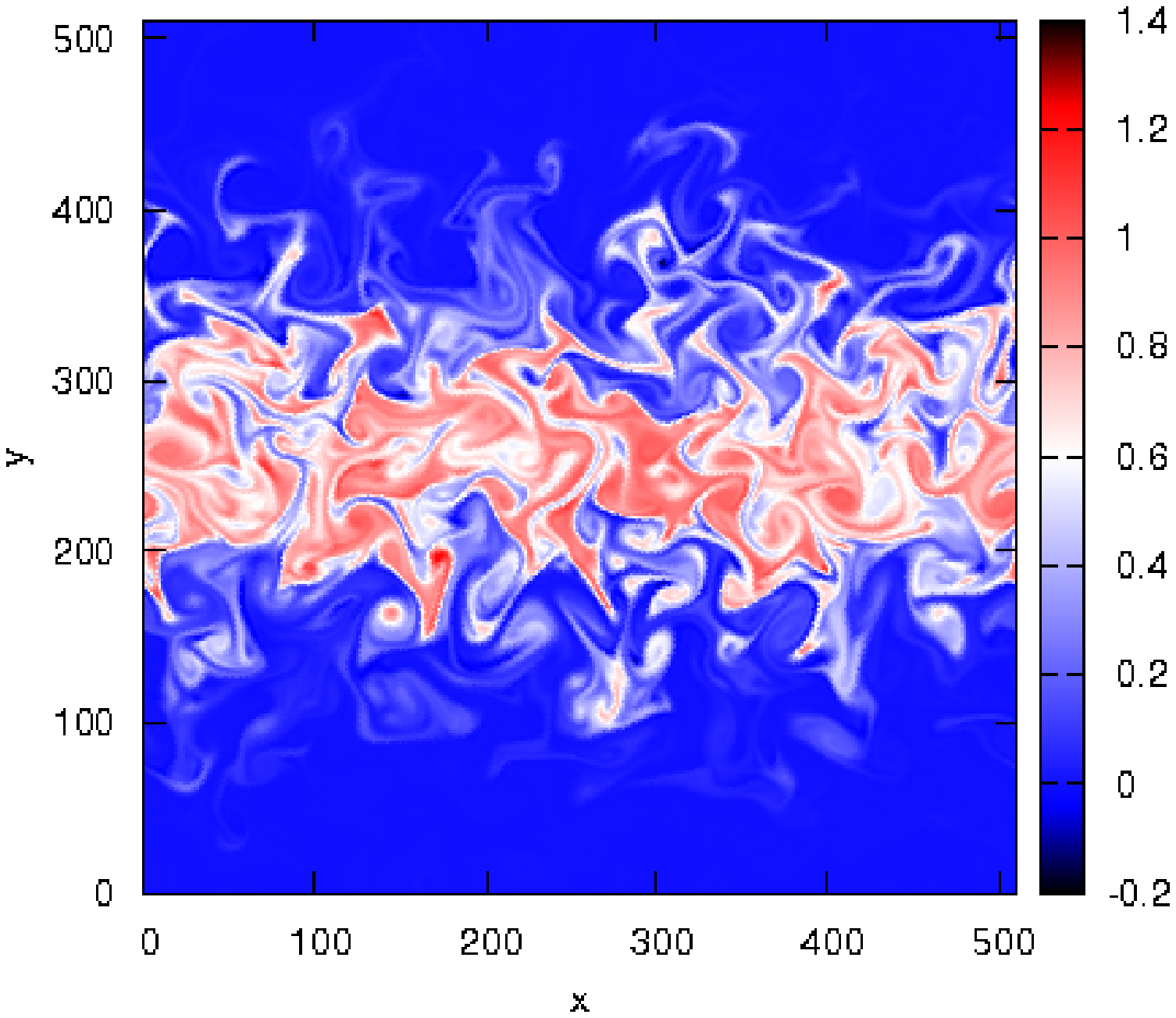}
\includegraphics[width=8.0cm]{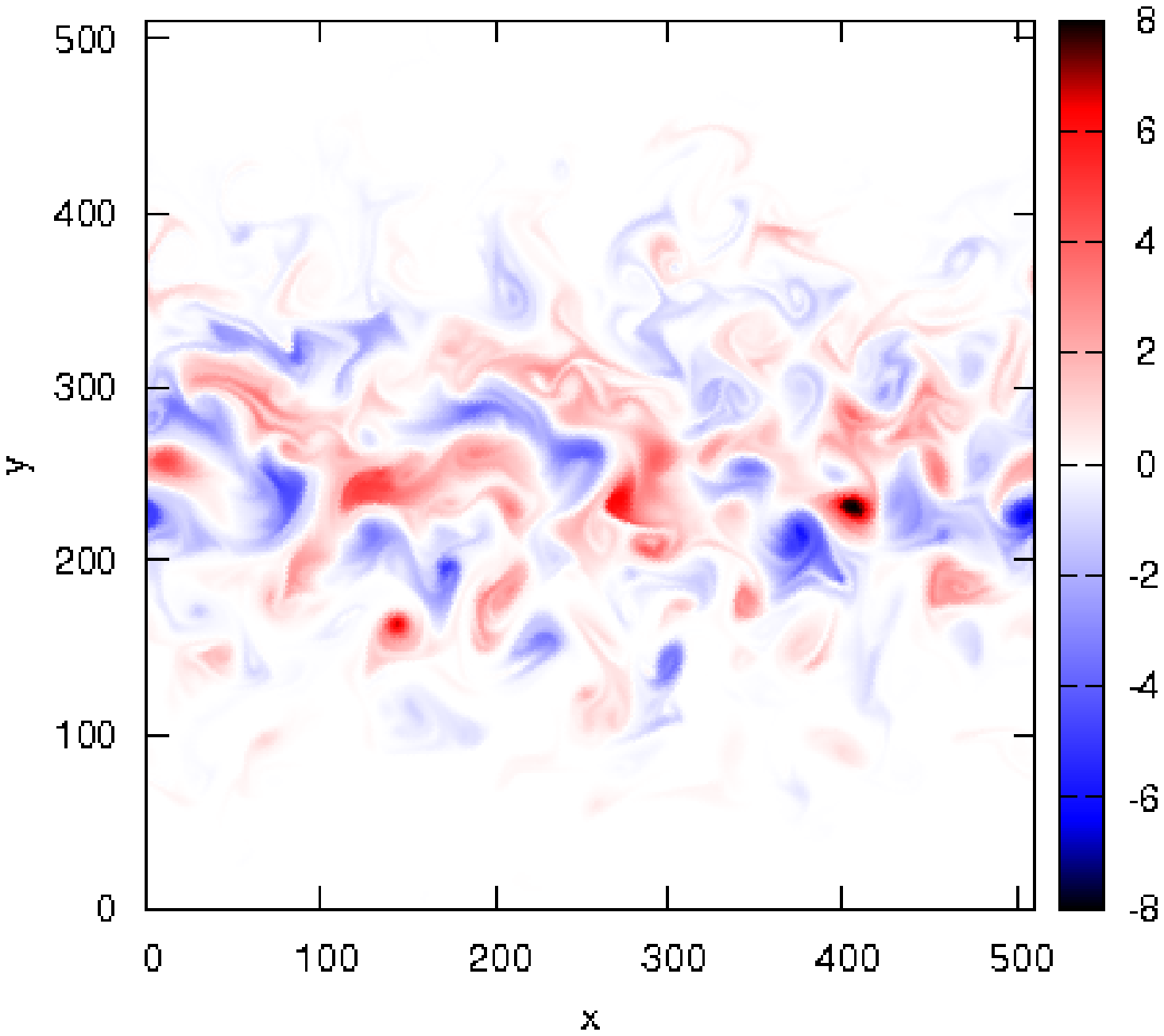}
\caption{ \sl
A Gaussian trace ion puff placed in the turbulent plasma is initially
localized around the center of the $y$ domain (and constant in $x$).
Left: impurity density $n_z(x,y)$. Right: the product function
$\Omega(x,y)\cdot n_z(x,y)$ shows that in this case no preferential sign of alignment
between impurity density and vorticity is found.}
\label{f:fig3}
\end{figure}

The situation is different if there is no background distribution of the trace
ion species, but only a smaller localized cloud (diffusing over time) with an
initial spatial extension in the same order of magnitude as the turbulence scales.
Then the cloud can have global gradients of its density in all
directions with respect to the background electron (and primary ion) gradient,
and the signs of $r$ approximately can cancel to zero by integration over the
computational domain, while the absolute correlation coefficient $|r|$ remains
near unity. 

The turbulent spreading of such a initially Gaussian trace ion
puff (with zero background) strongly localized in $x$ or $y$ is shown in
Figs.~\ref{f:fig2} and \ref{f:fig3}, respectively.
In Fig.~\ref{f:fig2} a Gaussian trace ion puff placed in the turbulent plasma is initially
localized around the center of the $x$ domain and constant in $y$. 
The product function $\Omega(x,y)\cdot n_z(x,y)$ shows (in the left part of the 
figure) in comparison with the impurity gyrocenter density $n_z(x,y)$ (right figure) that
$n_z(x,y)$ and $\Omega(x,y)$ are preferentially negatively aligned in the
right half of the domain, where the initial density gradients are co-aligned,
and positively aligned on the left. This behaviour is in accordance with the
previous result for a constant background gradient, as has been shown in
Fig.~\ref{f:fig1} 
For the $x$-symmetric case in Fig.~\ref{f:fig3} no preferential sign of alignment
between impurity gyrocenter density and vorticity is found.

Also this case the perturbed fluid particle density $N_z$ is still correlated
with vorticity according to the relation $D_t N_z = \mu_z D_t \Omega$. 
The preceding discussion of the present results was funded within the
gyrocenter density $n_z$ representation of the gyrofluid model.
It remains to be clarified when the advection by the background gradient or
the inertial contribution is the dominant factor that determines alignment of the
fluid particle density $N_z = n_z + \mu_z \Omega$ with vorticity.

\begin{figure} 
\includegraphics[width=9.0cm]{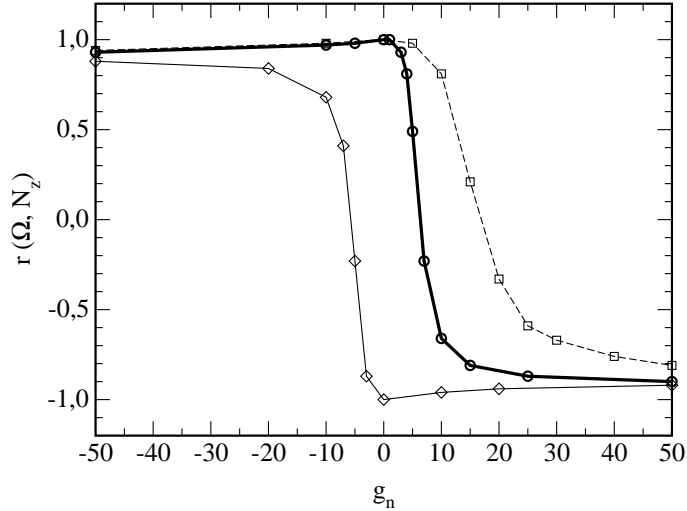}
\caption{\sl Correlation coefficient $r(\Omega,N_z)$ as a function of the impurity
  density parameter $g_n$ for $\mu_z=10$ (circles, bold solid line),
  $\mu_z=30$ (squares, thin dashed line) and $\mu_z=-10$ (diamonds, thin solid
  line): the alignment of impurity particle density with vorticity changes
  sign around $g_n \approx \mu_z/2$.  
}
\label{f:fig4align}
\end{figure}

In Fig.~\ref{f:fig4align} the correlation coefficient $r(\Omega,N_z)$ is shown as
a function of the impurity density gradient parameter $g_n$ at $d=2$ for
$\mu_z=10$  (circles, bold solid line), $\mu_z=30$ (squares, thin dashed
line) and $\mu_z=-10$ (diamonds, thin solid line). 
The alignment of perturbed impurity particle density with vorticity
changes sign around $g_n \approx \mu_z/2$: for $g_n \ll \mu_z$ the alignment
directly follows the vorticity with $r(\Omega,N_z)\approx 1$, and for $g_n \gg
\mu_z$ the alignment is reversed towards negative vorticity with
$r(\Omega,N_z) \approx -1$. 

The computations have been repeated for a non-adiabatic parallel coupling
coefficient $d=0.1$, for which the results remain very similar. The
overall impression given by Fig.~\ref{f:fig4align} does not change
signficantly in this case, except for slight (order of per cent) differences
within fluctuation averaging errors.

The sign of alignment is thus determined by the strength of the impurity
gradient in relation to the mass-to-charge ratio of the impurities. 
For example, for singly positively charged heavy molecules (like typical
space biomolecules) with $\mu_z \gg 1$ and small impurity gradient $g_z
\approx 1$ the alignment is tendentially towards positive vorticity.
For vanishing impurity gradient ($g_z=0$) the correlation $r(\Omega,N_z)$ is
always exactly $+1$ for positively charged impurities, and $-1$ for negative impurities.

The basic conclusion is that for given background gradients and magnetic field
direction the perturbed trace ion species gyrocenter density dynamically
aligns with a definite sign of the electron density fluctuations, and thus of
the electrostatic potential fluctations, and consequently of vorticity ${\bf
  \Omega}$: for example, an excess of trace ions aggregates 
within vortices of clockwise direction, and a deficit is found in vortices
with anti-clockwise direction (or vice versa, depending on global
parameters). 

While, as usual in a fully developed turbulent state, vortices of
both signs appear equally likely and evenly distributed over all turbulent
scales, the trace particle aggregation on drift scales emerges with one prefered
rotationality with respect to the background magnetic field and background
particle gradients. 

\section{Chiral molecular aggregation in drift wave turbulence}

Finally, a possible relevance of the asymmetric alignment effect on molecular
chemistry in a magnetized space plasma environment is suggested.

Homochirality of biomolecules -- the fact that the essential chemical building
blocks of life have a certain handedness while synthetic production leads to
equal (racemic) distribution of left-handed and right-handed chiral structures
-- has ever since its discovery by Pasteur \cite{Pas48} posed a formidable puzzle
\cite{Mas84,Lou02}. Pasteur already had unsuccessfully tried to identify physical
causes for this biological symmetry breaking, for example by imposing
chirality through fluid vortices in a centrifuge, and by exposing chemical
solutions to a magnetic field \cite{Pas84}. 

A number of theories and experiments on the origin of chirality have since
been put forward, like effects of circularly polarized light on the molecular
reactions \cite{Huc96,Bai98,Bow01} or a combination of magnetic field and
non-polarized light \cite{Rik00}, or possible electroweak effects on quantum
chemistry \cite{Ber01}. All of these mechanisms could be active in interplanetary and
interstellar space, and would hint on an extraterrestrial origin of early
fundamental biomolecules. 

That chirality can in principle indeed be imposed by rotational forces has
been confirmed experimentally for different situations \cite{Rib01,Mic12}, but what 
mechanism could invoke a specific directionality in turbulent (terrestrial or
space) fluids or plasmas, where vortices of both senses of rotation
generically occur mixed across all scales, has been left an open
question. Here we argue that rotation asymmetric ion aggregation in drift
vortices in magnetized space plasmas constitutes a mechanism for fostering a
truly chiral environment for enantiomeric selective extraterrestrial formation
of biomolecules. 

The drift scaling conditions and possibilities for occurrence of drift wave
turbulence are well fulfilled for a number of typical space plasma parameters
and magnetic field strengths, like in (warm ionized) clouds in the interstellar medium
\cite{How06}.  

The background pressure or density gradient length of molecular ion species in
the interstellar medium can take values across many orders of magnitude. While
interstellar clouds range in size between a few and hundreds of parsecs (1 pc
= $3.0857 \cdot 10^{16}$~m) and thus have global gradient lengths of the same order, the
local gradients can be set by macroscopic turbulence (resulting from large
scale magnetohydrodynamic motion or by chaotic external drive through winds
and jets) and vary widely, from observable scales of 100 pc down to 1000 km \cite{Gae11}. 

Then again, the gradient of molecular ion density can also be set by a
spatially varying degree of ionization, for example through inhomogeneous
irradiation at the edge of molecular clouds. Magnetic field strengths in the
interstellar medium can occur up to a few $\mu$G to mG, and temperatures range
between 10 K in cold molecular clouds to 10000 K in warm ionized interstellar
medium. The resulting drift (and vortex) scale $\rho_s$ is in the order
between a few hundred meters and a few hundred kilometers. 

Typical vortex life times are in the order between seconds and hours.
The mean free path (parallel to a magnetic field) between particle collisions
is in the order of 1000 km, so that the plasma is only weakly collisional. 
The drift ratio is in the maximum  order of $\delta = \rho_s / L_p = 0.1$ and
smaller, thus well below unity.  

The relative fluctuation amplitudes (compared to background values of the plasma density)
in drift wave turbulence are typically in the order of the drift ratio and
thus here in the range of a few percent or below. Compared to macroscopic
flow-driven or magnetohydrodynamic (MHD) turbulence, drift wave
turbulence is most effective on smaller scales, in the order of $\rho_s$, and with
relatively small amplitude. 
Drift wave vortices in the size between a few hundred meters and hundreds of
kilometers are thus expected to be present in most interstellar plasmas. 

Next we consider, if such vortices truly can account for a chiral physical
effect. Chiral synthesis usually requires a truly chiral influence
\cite{Bar86,Bar12}: the parity
transformation $(P: x, v \rightarrow -x, -v)$ has to result in a mirror asymmetric state
that needs to be different from simply a time reversal $(T: t \rightarrow -t)$ plus a
subsequent rotation by $\pi$. A purely 2D vortex can not exert a truly chiral
influence, while 3D funnel-like fluid vortices in principle do. While a bulk
rotation per se can not cause any direct polarizing effect on the reaction
path for molecular synthesis \cite{Fer99}, particle aggregation on a supramolecular
level in vortex motion has been experimentally shown to be able to lead to
chiral selection \cite{Rib01}. To date it however remains unclear how chirality from a
supramolecular aggregate (clusters or dust) could be transfered to single
molecules \cite{Fer01}. 

In the case of drift wave turbulence the parallel wave-like dynamics imposes a
wave vector $k_{||}$ that under $P$ changes direction with respect to the
vorticity ${\bf \Omega}$ and ${\bf B}$ (which themselves are pseudovectors and
remain invariant under $P$), equivalently to a 3D vortex tube. Drift wave
turbulence is thus truly chiral, although it characteristically appears
quasi-two-dimensional. In case of a large-scale density gradient $\nabla_{||}
n$ along ${\bf B}$ the (small) $k_{||}$ is given by this gradient length,
otherwise it is determined by local fluctuations. Molecular ions are advected
by the drift vortex motion, and are subjected to chiral aggregation onto
neutral molecules, clusters or dust particles that are not participating in
the plasma rotation.   

If in addition (as to be rather expected in most cases) a trace ion background
density gradient in the direction parallel to ${\bf B}$ is present, the resulting
parallel wave vector $k_{||}$ prescribes in combination with 
vorticity ${\bf \Omega}$ a definite true chiral vortex effect. Chemical reactions in this
system occur by collisions or aggregation of ions, that participate
in the chiral vortex rotation, with neutral molecules or clusters of the
quiescent (non rotating) background gas. The co- or counter alignment of ${\bf B}$
with  ${\bf \Omega}$ can further act as chiral catalyst. 

The chiral influence however changes sign in different sides or regions of the
(large-scale) molecular cloud, when the relative direction of background
particle gradient and magnetic field direction reverses. This implies that
different regions of the interstellar medium favour different tendencies for
chiral selective aggregation, and thus for a potential enantiomeric excess in
molecular syntheses. The excess rate could be expected in the order of the relative
density fluctuations in drift wave turbulence, of a few percent. 

A possible problem that could not be addressed with the present electrostatic
model is whether chiral alignment may be broken by strong Alfv\`enic activity. 
Electromagnetic computations for moderate beta values (a few percent as for
fusion edge plasmas) in ref.~\cite{Sco05} for absolute correlation have
however shown no significant deviation of the alignment character compared to
electrostatic computations, so that chiral alignment may be expected to
survive for finite beta.
 
\bigskip

Chiral aggregation in drift vortices should therefore be feasible locally
at least in lower beta regions of the interstellar or interplanetary media. 
The necessary conditions for chiral aggregation thus appear rather restrictive
and cosmologically rare. 
It may be concluded that either the suggested mechanism for chiral synthesis is
subdominant (compared to any other of proposed or yet unknown mechanisms), or
that a chiral excess of molecules should cosmologically be a rather rare
phenomenon itself.


\section*{Acknowledgements}

The author thanks B.D.~Scott (IPP Garching) for valuable discussions on
gyrofluid theory and computation.
This work has been funded by the Austrian Science Fund (FWF) Y398.
	 

\section*{Appendix}

The correspondence between the present cold-ion gyrofluid model with linear
polarization and the HW model of Priego et al. \cite{Pri05} including inertial
nonlinear polarization effects is in the following briefly lined out. 

For a general discussion on nonlinear polarization and dissipative
correspondence between low-frequency fluid and gyrofluid equations we refer to
ref.~\cite{Sco07} by Scott.

The Priego model (with notations and normalizations adopted to fit ours)
consists of the 2D Hasegawa-Wakatani (HW) model
\begin{eqnarray}
& & D_t (N_e-x) = d (\phi -N_e) \\
& & D_t \Omega = d(\phi-N_e)
\label{e:hw}
\end{eqnarray}
where $\Omega = \nabla^2 \phi$ is the vorticity, $D_t = \partial_t + {\bf v}_E \cdot
\bnabla$ is the convective derivative with ${\bf v}_E = ({\bf B}/|B|) \times
\bnabla \phi$, and $D_t x = - \partial_y \phi$ in a local model where the
background density gradient enters into the length scale normalisation.
For simplicity we neglect in this presentation all dissipation terms (where
Priego et al. have used normal second order dissipation, while we used fourth
order hyperviscous dissipation terms). 
Here capital letters are used for the fluid particle densities $N_s$ for
distinction to the gyrocenter densities $n_s$ of the gyrofluid model.
The particle densities fulfill the direct quasi-neutrality condition $N_e =
N_i$ for $N_z \ll N_{i0}$.

The massive trace impurities are in Priego's model assumed to be passively
advected, but due to their inertia respond to the velocity ${\bf v}_z = {\bf v}_E + {\bf
  v}_{pz}$ with the additional polarization drift velocity ${\bf v}_{pz} = - \mu_z
D_t \nabla \phi$ in the global compressional impurity continuity equation
\begin{equation}
\partial_t N_z + \bnabla \cdot ( N_z {\bf v}_z) = 0
\end{equation}
for the full impurity density $N_z$ (whereas only fluctuations of $N_e$
are evolved in the HW equation).
For incompressional ExB velocity (when the magnetic field is homogeneous) this
is equivalent to the full (global) impurity density equation used by Priego et
al., which is (in our notation) given in ref.~\cite{Pri05} as 
\begin{equation}
D_t N_z - \bnabla \cdot ( \zeta N_z D_t \bnabla \phi) = 0
\label{e:priego}
\end{equation}
with $\zeta \equiv \mu_z \delta_0$. The factor $\delta_0 = \rho_s / L_{ne}$
stems from the normalization $\phi \rightarrow \delta_0^{-1} e \tilde
\phi / T_{e0}$ of the local HW model. 

This introduces a nonlinear polarization term into the dynamics of 
massive impurities. 
In a local approximation (corresponding to the assumption of small
fluctuations on a large static background), where only the ExB advection is
kept as the sole nonlinearity, this equation reduces to:
\begin{equation}
D_t N_z = \zeta N_{z0} D_t \Omega.
\label{e:priegolin}
\end{equation}

Now we derive a set of HW-like vorticity-density fluid equations from the
(cold ion) gyrofluid model, and compare this with the Priego model.

The present gyrofluid model consists of the local gyrocenter density equations
\begin{eqnarray}
D_t (n_e + g_e x) &=& d(\phi - n_e) \\
D_t (n_i + g_i x) &=& 0 \\
D_t (n_z + g_z x )&=& 0 
\end{eqnarray}
and the local quasi-neutral polarization equation
\begin{equation}
\sum_s \left[ {a_s \over \tau_s} (\Gamma_0 -1) \phi + a_s \Gamma_1 n_s \right] = 0.
\end{equation}
where  $\tau_s = T_{s0}/T_{e0}$, $a_s = Z n_s/ n_e$ and $\mu_s = m_s/(Z m_i)$. 
The gyrofluid advective derivative $D_t n_s = \partial_t n_s + [\psi,n_s]$
includes the gyro-averaged potential $\psi = \Gamma_1 \phi$ 
with $\Gamma_0 = [1+b]^{-1}$
and $\Gamma_1 = \Gamma_0^{1/2} = [1+(1/2)b]^{-1}$ 
in Pade approximation, where $b = -\tau_s \mu_s \nabla^2$. 
In Taylor approximation one would have $\Gamma_0 \approx [1-b]$ and $\Gamma_1 \approx
[1-(1/2)b]$. The species coefficients are  $\tau_e=1$, $a_e = -1$, $\mu_e=0$
for electrons, and $a_i=1$ for ions.  

The background gradient terms $g_s = \partial_x \ln n_{s0}$ fulfill (for zero
background vorticity) the global quasi-neutrality condition $g_e = g_i + a_z
g_z \approx g_i$ when $g_i \ll 1$ for trace impurities. In the normalization
of perpendicular length scales by $L_{\perp} = L_n = |\partial_x \ln n_{e0}|^{-1}$
we then have $g_e=g_i=1$.

The perturbed particle and gyrocenter densities are connected via the elements of the
polarization equation as \cite{Sco07} 
\begin{equation}
a_s N_s = a_s \Gamma_1 n_s + \mu_s a_s \nabla^2 \phi
\end{equation}
so that gyrofluid polarisation corresponds to the fluid particle
quasi-neutrality condition $\sum_s a_s N_s = 0$. 

From now on again cold ions with $\tau_i=\tau_z=0$ are assumed, and thus
$\psi=\phi$ and $\Gamma_0 = \Gamma_1 = 1$. The local polarization then reduces to 
$\rho_m \nabla^2 \phi =  n_e - n_i -a_z n_z$ with $\rho_m = \sum_s a_s \mu_s = 1+ a_z \mu_z$.

It can be immediately seen that the local cold-ion gyrofluid model already includes the
(linear) polarization effect that has been added in the Priego model, when the
gyrocenter density in the impurity continuity equation $D_t n_z =0$ is
substituted by the fluid particle density: 
\begin{equation}
D_t N_z = D_t \alpha_z n_z + D_t \mu_z \alpha_z \Omega = \mu_z \alpha_z D_t \Omega
\end{equation}
where $\alpha_s = a_s/Z = n_{s0}/n_{e0}$.
This is identical to the linearized form eq.~\ref{e:priegolin} of the Priego
model (where the full impurity density is evolved so that on the right hand
side $\mu_z \alpha_z \rightarrow \mu_z \delta_0 N_{z0}$). 
Completely passive trace ions would be achieved in the gyrofluid model for
$\mu_z=0$ so that $\rho_m=1$.

A fully global set of nonlinear gyrofluid equations 
would include the nonlinear quasi-neutral polarization equation \cite{Str04},
which (with $M_s=m_s/m_i$) is given as: 
\begin{equation}
\sum_s \bnabla \cdot (M_s n_s \bnabla \phi) + \sum_s q_s \Gamma_1 n_s = 0.
\label{e:globalpol}
\end{equation}
This equation is expensive to solve numerically and requires e.g. the use of a
multi-grid solver. Many codes therefore apply the linearized form, which
can be treated by standard fast Poisson solvers. This is usually regarded as
a valid approximation as long as the turbulent density fluctuations are small
compared to the background plasma density. 

The trace of the (cold ion) global polarization eq.~\ref{e:globalpol} again
delivers the fluid-gyrofluid density relations:
\begin{equation}
N_s = n_s + \bnabla \cdot (M_s n_s \bnabla \phi).
\end{equation}
From this we get the fluid impurity continuity equation
\begin{equation}
D_t N_z = D_t \bnabla \cdot (M_z n_z \bnabla \phi) = \bnabla \cdot (M_z
n_z D_t \bnabla \phi)
\end{equation}
where twice $D_t n_z = 0$ has been used. The term on the right hand side
however still contains the gyrofluid density $n_z$. The approximation
\begin{equation}
D_t N_z \approx  \bnabla \cdot (\zeta N_z D_t \bnabla \phi)
\label{e:result}
\end{equation}
with the fluid particle density $N_z$ appears acceptable if either $\zeta \ll
1$ or if only small turbulent fluctuations on a static background can be assumed, 
so that the additional gyrocenter correction could be considered an order
smaller than the other terms.
The resulting fluid impurity equation~\ref{e:result} is then equivalent to
  eq.~\ref{e:priego} as used by Priego et al.~\cite{Pri05}.

Summing up, we here retain the effects of a co- or counter-aligned
impurity density background gradient and linear polarization, but
neglect nonlinear polarization effects. 
Priego et al. have considered a constant impurity background density, but
due to their inclusion of a nonlinear polarization term were able to also treat
additional nonlinear clustering and aggregation effects of impurities by
inertia within vortices.

\end{document}